\documentclass[twoside]{article}
\usepackage{amsmath, graphicx, amssymb, booktabs, multirow, makecell, hyperref, qic}
\usepackage[caption=false]{subfig}

\newcommand{\braket}[2]{{\left\langle #1 \middle| #2 \right\rangle}}

\newcommand{\ket}[1]{{\left| #1 \right\rangle}}
\newcommand{\ketbra}[2]{{\left| #1 \middle\rangle \middle \langle #2 \right|}}
\newcommand{\fref}[1]{Fig.~\ref{#1}}

\newcommand{\sref}[1]{Section~\ref{#1}}

\newcommand{\ie}[1]{{\textit{i.e.}},}

\begin{document}

\setlength{\textheight}{8.0truein}    

\runninghead{Conserved Quantities in Linear and Nonlinear Quantum Search}
            {D.~A.~Meyer and T.~G.~Wong}

\normalsize\textlineskip
\thispagestyle{empty}
\setcounter{page}{1}



\alphfootnote

\fpage{1}

\centerline{\bf CONSERVED QUANTITIES IN LINEAR AND}
\vspace*{0.035truein}
\centerline{\bf NONLINEAR QUANTUM SEARCH}
\vspace*{0.37truein}
\centerline{\footnotesize DAVID A.~MEYER}
\vspace*{0.015truein}
\centerline{\footnotesize\it Department of Mathematics, University of California, San Diego}
\baselineskip=10pt
\centerline{\footnotesize\it 9500 Gilman Drive, La Jolla, California 92093, USA}
\vspace*{10pt}
\centerline{\footnotesize THOMAS G.~WONG}
\vspace*{0.015truein}
\centerline{\footnotesize\it Department of Physics, Creighton University}
\baselineskip=10pt
\centerline{\footnotesize\it 2500 California Plaza, Omaha, Nebraska 68178, USA}

\vspace*{0.21truein}

\abstracts{In this tutorial, which contains some original results, we bridge the fields of quantum computing algorithms, conservation laws, and many-body quantum systems by examining three algorithms for searching an unordered database of size $N$ using a continuous-time quantum walk, which is the quantum analogue of a continuous-time random walk. The first algorithm uses a linear quantum walk, and we apply elementary calculus to show that the success probability of the algorithm reaches 1 when the jumping rate of the walk takes some critical value. We show that the expected value of its Hamiltonian $H_0$ is conserved. The second algorithm uses a nonlinear quantum walk with effective Hamiltonian $H(t) = H_0 + \lambda|\psi|^2$, which arises in the Gross-Pitaevskii equation describing Bose-Einstein condensates. When the interactions between the bosons are repulsive, $\lambda > 0$, and there exists a range of fixed jumping rates such that the success probability reaches 1 with the same asymptotic runtime of the linear algorithm, but with a larger multiplicative constant. Rather than the effective Hamiltonian, we show that the expected value of $H_0 + \frac{1}{2} \lambda|\psi|^2$ is conserved. The third algorithm utilizes attractive interactions, corresponding to $\lambda < 0$. In this case there is a time-varying critical function for the jumping rate $\gamma_c(t)$ that causes the success probability to reach 1 more quickly than in the other two algorithms, and we show that the expected value of $H(t)/[\gamma_c(t) N]$ is conserved.}{}{}

\vspace*{10pt}


\vspace*{1pt}\textlineskip


\section{Introduction}

Conserved quantities play important roles in physics: energy, momentum, and angular momentum are conserved in systems that are time, translation, and rotation invariant, respectively. In quantum mechanics, the wave function $\psi$ of a system with Hamiltonian $H_0$ evolves according to Schr\"odinger's equation,
\begin{equation}
    \label{eq:SE}
    i \hbar \frac{\partial \psi}{\partial t} = H_0 \psi,
\end{equation}
where we will work in units where the reduced Planck constant $\hbar = 1$ throughout this paper. The Hamiltonian $H_0$ must be Hermitian (\textit{i.e.}, self-adjoint), which causes the time evolution to be unitary. Then, $|\psi|^2 = \braket{\psi}{\psi}$ is a conserved quantity, meaning a normalized wave function stays normalized, which is necessary for the probabilistic interpretation of the wave function \textit{\`a la} the Born rule. This may not be the only conserved quantity, however. For example, if the Hamiltonian is time-independent, then the expected value of the Hamiltonian $\langle H_0 \rangle = \langle \psi | H_0 | \psi \rangle$, \textit{i.e.}, the average energy of an ensemble of identically-prepared systems, is also conserved \cite{Griffiths2018}. The time-independence of the Hamiltonian and the conservation of its expected value hold for many quantum algorithms based on continuous-time quantum walks \cite{CCDFGS2003,CG2004,FGG2008}, where the particle walks on the vertices of a graph, and the kinetic energy of the system is proportional to the discrete Laplacian. For overviews of quantum walks, see \cite{Kempe2003,Ambainis2003,Venegas2012,DiMolfetta2024}. In \sref{section:linear}, we will give an example of this by reviewing how a continuous-time quantum walk searches the complete graph for a marked vertex \cite{CG2004,Wong35}, which is the combinatorial formulation of the unstructured search problem of Grover's algorithm \cite{Grover1996}. In the process, we will give a new derivation of the evolution of the algorithm for arbitrary jumping rates, as well as give a new derivation---using only elementary calculus---of the critical jumping rate that causes the algorithm to succeed with certainty. We will explicitly show that for the Hamiltonian used in the search algorithm, $\langle H_0 \rangle$ is conserved.

Some many-body quantum systems are approximately described by an effectively nonlinear Schr\"odinger-type equation with a cubic nonlinearity proportional to $|\psi|^2\psi$, \textit{i.e.},
\begin{equation}
    \label{eq:NLSE}
    i \hbar \frac{\partial \psi}{\partial t} = \left( H_0 + \lambda |\psi|^2 \right) \psi,
\end{equation}
where the effective Hamiltonian
\begin{equation}
    \label{eq:H(t)}
    H(t) = H_0 + \lambda |\psi|^2
\end{equation}
depends on time because it depends on the state $\psi$, which evolves with time. A Bose-Einstein condensate \cite{Bose1924,Einstein1924,Einstein1925} is an example of such a system, with the Gross-Pitaevskii equation \cite{Gross1961,Pitaevskii1961} describing it in the mean-field limit and taking the form of the cubic nonlinear Schr\"odinger equation \eqref{eq:NLSE}; for reviews, see \cite{Dalfovo1999,Morsch2006}. As before, $|\psi|^2 = 1$ is conserved, but since $H(t)$ is time-dependent, $\langle H(t) \rangle = \langle \psi | H(t) | \psi \rangle$ is not conserved. When $H_0$ is a constant, however, then $\langle H_0 + \frac{1}{2} \lambda|\psi|^2 \rangle = \langle \psi | H_0 + \frac{1}{2} \lambda |\psi|^2 | \psi \rangle$ is conserved \cite{Rogel-Salazar2013}. In \sref{section:repulsive}, we will give an example of this by reviewing how a nonlinear quantum walk with $\lambda > 0$, which corresponds to a Bose-Einstein condensate with repulsive interactions, searches the complete graph for a marked vertex \cite{Kahou2013}, and we will explicitly show that $\langle H_0 + \frac{1}{2} \lambda|\psi|^2 \rangle$ is conserved for the algorithm.

If $H_0$ depends on time, however, then $\langle H_0 + \frac{1}{2} \lambda|\psi|^2 \rangle$ is generally not conserved. In \sref{section:attractive}, we give an example of this by reviewing how a nonlinear quantum walk with $\lambda < 0$, which corresponds to a Bose-Einstein condensate with attractive interactions, searches a complete graph of $N$ vertices for a marked vertex \cite{Wong3}. In this algorithm the jumping rate of the quantum walk is a critical, time-dependent function $\gamma_c(t)$, which causes $H_0$ to be time-dependent as well. While $\langle H_0 + \frac{1}{2} \lambda|\psi|^2 \rangle$ is not conserved in this case, we show that the critical jumping rate is such that $\langle H_0 / [\gamma_c(t) N] \rangle$ is conserved instead.

Thus, in this tutorial, we will review three continuous-time quantum walk algorithms for searching the complete graph: a linear algorithm in \sref{section:linear}, a nonlinear algorithm corresponding to a Bose-Einstein condensate with repulsive interactions in \sref{section:repulsive}, and a nonlinear algorithm corresponding to a Bose-Einstein condensate with attractive interactions in \sref{section:attractive}. For each, we will demonstrate a conserved quantity related to the (effective) Hamiltonian of the system, thus making connections between quantum algorithms, conservation laws, and many-body quantum systems. We conclude in \sref{section:conclusion}.


\section{\label{section:linear}Linear Quantum Search}

\begin{figure}
\begin{center}
    \includegraphics{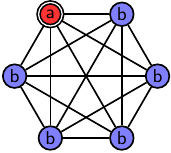}
	\vspace*{13pt}
    \fcaption{\label{fig:complete}A complete graph of $N = 6$ vertices. A vertex is marked, as indicated by a double circle. In the search algorithm, vertices that evolve identically have the same label and color.}
\end{center}
\end{figure}

\begin{figure}
\begin{center}
    \subfloat[] {
        \label{fig:linear-n100-a}
        \includegraphics{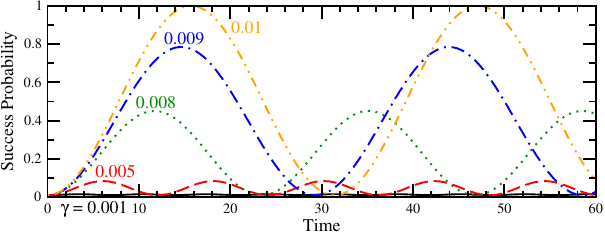}
    }
    
    \subfloat[] {
        \label{fig:linear-n100-b}
        \includegraphics{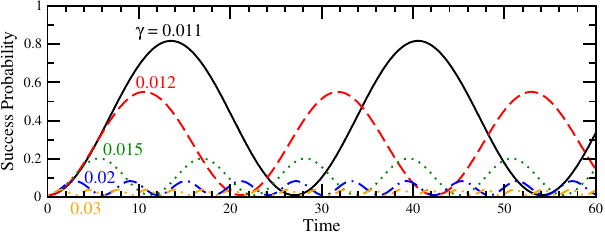}
    }
	\vspace*{13pt}
    \fcaption{\label{fig:linear-n100}Success probability versus time for (linear) quantum search on the complete graph with $N = 100$ vertices and various jumping rates. In (a), the solid black curve is $\gamma = 0.001$, the dashed red curve is $\gamma = 0.005$, the dotted green curve is $\gamma = 0.008$, the dot-dashed blue curve is $\gamma = 0.009$, and the dot-dot-dashed orange curve is $\gamma = 0.01$. In (b), the solid black curve is $\gamma = 0.011$, the dashed red curve is $\gamma = 0.012$, the dotted green curve is $\gamma = 0.015$, the dot-dashed blue curve is $\gamma = 0.02$, and the dot-dot-dashed orange curve is $\gamma = 0.03$.}
\end{center}
\end{figure}

In quantum mechanics, the Hamiltonian $H_0$ is the operator corresponding to the total energy of the system, which consists of a kinetic energy term that is proportional to Laplace's operator $\nabla^2$ and a potential energy term $V$ \cite{Griffiths2018}:
\[ H_0 = -\frac{\hbar^2}{2m} \nabla^2 + V. \]
For a quantum walk on a graph of $N$ vertices, the vertices label computational basis states $\ket{1}, \ket{2}, \dots, \ket{N}$, and in general, the state is a superposition or complex linear combination over the vertices. Then, Laplace's operator in the continuum should be replaced by the discrete Laplacian $L$:
\[ H_0 = -\gamma L + V, \]
where $L$ is an $N \times N$ matrix with $L_{ij} = 1$ if vertices $i$ and $j$ are adjacent, $L_{ii} = -\deg(i)$ is the negative of the number of neighbors of vertex $i$, and $L_{ij} = 0$ otherwise. For the complete graph of $N$ vertices, an example of which is shown in \fref{fig:complete}, the discrete Laplacian has $1$ for each off-diagonal element and $-(N-1)$ on the diagonal. We have also written the coefficient of $L$ as a parameter $\gamma$, which corresponds to the jumping rate of the quantum walk. To construct a quantum search algorithm \cite{CG2004}, an oracle ``marks'' a particular vertex $\ket{a}$ that we want to find, which manifests as a potential energy term:
\begin{equation}
    \label{eq:H0}
    H_0 = -\gamma L - \ketbra{a}{a}.
\end{equation}
Initially, we have no knowledge of where the marked vertex might be, so we guess each vertex equally by beginning in a uniform superposition over all $N$ vertices:
\begin{equation}
    \label{eq:psi(0)}
    \ket{\psi(0)} = \frac{1}{\sqrt{N}} \sum_{i = 1}^N \ket{i}.
\end{equation}
Then, the system evolves by Schr\"odinger's equation \eqref{eq:SE}, and since $H_0$ is time-independent,
\begin{equation}
    \label{eq:psi(t)}
    \ket{\psi(t)} = e^{-iH_0t} \ket{\psi(0)},
\end{equation}
where we have taken $\hbar = 1$. The success probability at time $t$ is the probability that measuring the position of the particle at time $t$ results in it being found at the marked vertex, so the success probability is
\begin{equation}
    \label{eq:p(t)}
    p(t) = \left| \braket{a}{\psi(t)} \right|^2.
\end{equation}
The evolution of this success probability is plotted in \fref{fig:linear-n100} for search on a complete graph of $N = 100$ vertices and with various values of the jumping rate $\gamma$. In \fref{fig:linear-n100-a}, the jumping rate starts off small as $\gamma = 0.001$ in the solid black curve, and the success probability is relatively unchanged from its initial value of $1/100 = 0.01$. As the jumping rate increases, however, the success probability evolves with a higher and higher peak, eventually reaching a peak success probability of $1$ when $\gamma = 0.01$. In \fref{fig:linear-n100-b}, $\gamma$ is increased further, and the peak success probability decreases, with larger $\gamma$ resulting in the success probability staying near its initial value.

To prove this behavior, we begin by noting that due to the symmetry of the problem, there are only two types of vertices: the marked vertex and the unmarked vertices, labeled in \fref{fig:complete} as $a$ and $b$. Vertices of the same type evolve identically, so the system evolves in a two-dimensional (2D) subspace spanned by the marked vertex $\ket{a}$ and the uniform superposition of unmarked vertices, which we will call $\ket{b}$:
\[ \ket{b} = \frac{1}{\sqrt{N-1}} \sum_{i\text{ unmarked}} \ket{i}. \]
In this basis, the initial state \eqref{eq:psi(0)} is
\begin{equation}
    \label{eq:psi(0)-2D}
    \ket{\psi(0)} = \frac{1}{\sqrt{N}} \ket{a} + \sqrt{\frac{N-1}{N}} \ket{b} = \frac{1}{\sqrt{N}} \begin{pmatrix} 1 \\ \sqrt{N-1} \end{pmatrix},
\end{equation}
and the Hamiltonian \eqref{eq:H0} is
\begin{equation}
    \label{eq:H0-2D}
    H_0 = \begin{pmatrix}
	   \gamma(N-1) - 1 & -\gamma \sqrt{N-1} \\
	   -\gamma \sqrt{N-1} & \gamma \\
    \end{pmatrix}.
\end{equation}
The (unnormalized) eigenvectors and eigenvalues of $H_0$ are
\begin{align*}
    &\psi_1 = \left( \gamma N - 2\gamma - 1 - \Delta E \right) \ket{a} - 2\gamma\sqrt{N-1} \ket{b}, && E_1 = \frac{\gamma N - 1 - \Delta E}{2}, \displaybreak[0] \\
    &\psi_2 = \left( \gamma N - 2\gamma - 1 + \Delta E \right) \ket{a} - 2\gamma\sqrt{N-1} \ket{b}, && E_2 = \frac{\gamma N - 1 + \Delta E}{2},
\end{align*}
where
\begin{equation}
    \label{eq:energy-gap}
    \Delta E = E_2 - E_1 = \sqrt{\gamma^2 N^2 - 2\gamma N + 4\gamma + 1}
\end{equation}
is the energy gap. Expressing the initial state \eqref{eq:psi(0)-2D} as a linear combination of these eigenvectors,
\[ \ket{\psi(0)} = \frac{-\gamma N + 1 - \Delta E}{4 \gamma \sqrt{N} \Delta E} \psi_1 + \frac{\gamma N - 1 - \Delta E}{4 \gamma \sqrt{N }\Delta E} \psi_2. \]
Then, using \eqref{eq:psi(t)}, the state of the system at time $t$ is 
\begin{align}
    \ket{\psi(t)}
        &= e^{-iH_0t} \left( \frac{-\gamma N + 1 - \Delta E}{4 \gamma \sqrt{N} \Delta E} \psi_1 + \frac{\gamma N - 1 - \Delta E}{4 \gamma \sqrt{N} \Delta E} \psi_2 \right) \nonumber \displaybreak[0] \\
        &= \frac{-\gamma N + 1 - \Delta E}{4 \gamma \sqrt{N} \Delta E} e^{-iE_1t} \psi_1 + \frac{\gamma N - 1 - \Delta E}{4 \gamma \sqrt{N} \Delta E} e^{-iE_2t} \psi_2 \nonumber \displaybreak[0] \\
        &= \frac{e^{-i(\gamma N - 1)t/2}}{4 \gamma \sqrt{N} \Delta E} \left[ \left( -\gamma N + 1 - \Delta E \right) e^{i \Delta E t/2} \psi_1 + \left( \gamma N - 1 - \Delta E \right) e^{-i \Delta E t/2} \psi_2 \right] \nonumber \displaybreak[0] \\
        &= \frac{e^{-i(\gamma N - 1)t/2}}{4 \gamma \sqrt{N} \Delta E} \Biggl\{ \left( -\gamma N + 1 - \Delta E \right) e^{i \Delta E t/2} \left[ \left( \gamma N - 2\gamma - 1 - \Delta E \right) \ket{a} - 2\gamma\sqrt{N-1} \ket{b} \right] \nonumber \\
            &\qquad\qquad\qquad+ \left( \gamma N - 1 - \Delta E \right) e^{-i \Delta E t/2} \left[ \left( \gamma N - 2\gamma - 1 + \Delta E \right) \ket{a} - 2\gamma\sqrt{N-1} \ket{b} \right] \Biggr\} \nonumber \displaybreak[0] \\
        &= \frac{e^{-i(\gamma N - 1)t/2}}{4 \gamma \sqrt{N} \Delta E} \Biggl\{ \biggl[ \Bigl( e^{i \Delta E t/2} - e^{-i \Delta E t/2} \Bigr) \Bigl( \left( -\gamma N + 1 \right) \left( \gamma N - 2\gamma - 1 \right) + \left( \Delta E \right)^2 \Bigr) \nonumber \\
            &\qquad\qquad\qquad\qquad + \Bigl( e^{i \Delta E t/2} + e^{-i \Delta E t/2} \Bigr) \Bigl( -\Delta E \left( \gamma N - 2\gamma - 1 \right) + \left( \gamma N - 1 \right) \Delta E \Bigr) \biggr] \ket{a} \nonumber \\
            &\qquad\qquad\qquad - 2\gamma\sqrt{N-1} \Biggl[ \left( e^{i \Delta E t/2} - e^{-i \Delta E t/2} \right) \left( -\gamma N + 1 \right) \nonumber \\
            &\qquad\qquad\qquad\qquad\qquad\qquad + \left( e^{i \Delta E t/2} +  e^{-i \Delta E t/2}\right) \left( -\Delta E \right) \Biggr] \ket{b} \Biggr\} \nonumber \displaybreak[0] \\
        &= \frac{e^{-i(\gamma N - 1)t/2}}{4 \gamma \sqrt{N} \Delta E} \Biggl\{ \biggl[ 2i\sin\left( \frac{\Delta E t}{2} \right) \left[ \left( -\gamma N + 1 \right) \left( \gamma N - 2\gamma - 1 \right) + \left( \Delta E \right)^2 \right] \nonumber \\
            &\qquad\qquad\qquad\qquad + 2\cos\left( \frac{\Delta E t}{2} \right) 2\gamma\Delta E \biggr] \ket{a} \nonumber \\
            &\qquad\qquad\qquad - 2\gamma\sqrt{N-1} \biggl[ 2i\sin\left( \frac{\Delta E t}{2} \right) \left( -\gamma N + 1 \right) \nonumber \\
            &\qquad\qquad\qquad\qquad\qquad\qquad + 2\cos\left( \frac{\Delta E t}{2} \right) \left( -\Delta E \right) \biggr] \ket{b} \Biggr\}. \label{eq:psi(t)-gamma}
\end{align}
Taking the norm-square of the amplitude of $\ket{a}$, the success probability \eqref{eq:p(t)} at time $t$ is
\begin{align}
    p(t)
        &= \frac{1}{16 \gamma^2 N \left( \Delta E \right)^2} \left\{ 4 \sin^2\left( \frac{\Delta E t}{2} \right) \left[ \left( -\gamma N + 1 \right) \left( \gamma N - 2\gamma - 1 \right) + \left( \Delta E \right)^2 \right]^2 \right. \nonumber \\
            &\qquad\qquad\qquad\qquad + \left. 16 \gamma^2 \left( \Delta E \right)^2 \cos^2\left( \frac{\Delta E t}{2} \right) \right\} \nonumber \\
        &= \left( \frac{\left( -\gamma N + 1 \right) \left( \gamma N - 2\gamma - 1 \right) + \left( \Delta E \right)^2}{2 \gamma \sqrt{N} \Delta E} \right)^2 \sin^2\left( \frac{\Delta E t}{2} \right) + \frac{1}{N} \cos^2\left( \frac{\Delta E t}{2} \right). \label{eq:p(t)-gamma}
\end{align}
This agrees with the curves in \fref{fig:linear-n100} with $N = 100$ and various values of $\gamma$. The success probability reaches a maximum value when the sine is 1 and the cosine is 0, \textit{i.e.}, at time
\begin{equation}
    \label{eq:t*-gamma}
    t_* = \frac{\pi}{\Delta E},
\end{equation}
at which the success probability peaks at
\begin{equation}
    \label{eq:p*-gamma}
    p_* =  \left( \frac{\left( -\gamma N + 1 \right) \left( \gamma N - 2\gamma - 1 \right) + \left( \Delta E \right)^2}{2 \gamma \sqrt{N} \Delta E} \right)^2.
\end{equation}

To the best of our knowledge, the above results for $\ket{\psi(t)}$ \eqref{eq:psi(t)-gamma}, $p(t)$ \eqref{eq:p(t)-gamma}, $t_*$ \eqref{eq:t*-gamma}, and $p_*$ \eqref{eq:p*-gamma} for arbitrary $\gamma$ are new, as previous literature \cite{Wong35} focused on the specific case when $\gamma = 1/N$, motivated by minimizing the energy gap \cite{CG2004} or by degenerate perturbation theory \cite{Wong5}. Now, we will provide another way to arrive at the $\gamma = 1/N$ case, namely using elementary calculus to prove that it is the value of $\gamma$ that maximizes the peak success probability $p_*$ \eqref{eq:p*-gamma}. Substituting the energy gap $\Delta E$ \eqref{eq:energy-gap} into $p_*$ \eqref{eq:p*-gamma} and differentiating the result with respect to $\gamma$, we get
\[ \frac{dp_*}{d\gamma} = \frac{4(N-1)(\gamma^2N^2 - 1)}{N(\gamma^2 N^2 - 2\gamma(N-1)+1)^2}. \]
Setting this equal to $0$ and solving for $\gamma$, the peak success probability $p_*$ is maximized when $\gamma$ takes a critical value of
\[ \gamma_c = \frac{1}{N}, \]
at which the energy gap \eqref{eq:energy-gap} is
\[ \Delta E_c = \frac{2}{\sqrt{N}}, \]
and the peak success probability \eqref{eq:p*-gamma} is
\[ p_{*c} = 1. \]
That is, $\gamma = \gamma_c = 1/N$ makes the algorithm deterministic, \textit{i.e.}, the marked vertex is found with certainty. Using \eqref{eq:t*-gamma}, this occurs at time
\[ t_{*c} = \frac{\pi}{2} \sqrt{N}, \]
which is the expected $O(\sqrt{N})$ scaling of Grover's algorithm \cite{Grover1996}. Furthermore, using \eqref{eq:p(t)-gamma}, the success probability evolves with time as
\[ p_c(t) = \sin^2 \left( \frac{t}{\sqrt{N}} \right) + \frac{1}{N} \cos^2 \left( \frac{t}{\sqrt{N}} \right), \]
which agrees with the dot-dot-dashed orange curve in \fref{fig:linear-n100-a}, where $N = 100$ and $\gamma = 1/100 = 0.01$.

Moving on to conserved quantities, since the Hamiltonian \eqref{eq:H0} is time-independent, its expected value is conserved, \textit{i.e.}, $\langle H \rangle = \langle \psi(t) | H | \psi(t) \rangle$ is constant in time \cite{Griffiths2018}. We will explicitly show this in the 2D subspace of our algorithm for pedagogical purposes, as the calculation for the nonlinear algorithm in the next section will use similar methods. To begin, we write the state of the system in the $\{ \ket{a}, \ket{b} \}$ basis as \begin{equation}
    \label{eq:psi(t)-2D-general}
    \ket{\psi(t)} = \alpha(t) \ket{a} + \beta(t) \ket{b} = \begin{pmatrix}
        \alpha(t) \\
        \beta(t) \\
    \end{pmatrix}
\end{equation}
for complex $\alpha$ and $\beta$, where for convenience, we have stopped writing the explicit time-dependence of $\alpha$ and $\beta$. The Hamiltonian \eqref{eq:H0-2D} takes the form
\begin{equation}
    \label{eq:H-2D-general}
    H_0 = \begin{pmatrix} a & b \\ b^* & d \end{pmatrix},
\end{equation}
where
\begin{equation}
    \label{eq:abd}
    a = \gamma(N-1) - 1, \quad b = -\gamma \sqrt{N-1}, \quad \text{and} \quad d = \gamma \\
\end{equation}
are real and time-independent, although we allow for $b$ to be complex for greater generality. Then, the expected value of $H_0$ is
\begin{align}
	\langle H_0 \rangle
		&= \langle \psi | H_0 | \psi \rangle \nonumber \\
		&= \begin{pmatrix} \alpha^* & \beta^* \end{pmatrix} \begin{pmatrix} a & b \\ b^* & d \end{pmatrix} \begin{pmatrix} \alpha \\ \beta \end{pmatrix} \nonumber \\
		&= \begin{pmatrix} \alpha^* & \beta^* \end{pmatrix} \begin{pmatrix} a\alpha + b\beta \\ b^*\alpha + d\beta \end{pmatrix} \nonumber \\
		&= a\alpha^*\alpha + b\alpha^*\beta + b^*\beta^*\alpha + d\beta^*\beta. \label{eq:expected-H}
\end{align}
We want to show that this is constant in time. Using the chain rule, its time-derivative is
\begin{align}
	\frac{d\langle H_0 \rangle}{dt} 
		&= \frac{\partial\langle H_0 \rangle}{\partial\alpha} \frac{d\alpha}{dt} + \frac{\partial\langle H_0 \rangle}{\partial\beta} \frac{d\beta}{dt} + \frac{\partial\langle H_0 \rangle}{\partial\alpha^*} \frac{d\alpha^*}{dt} + \frac{\partial\langle H_0 \rangle}{\partial\beta^*} \frac{d\beta^*}{dt} \nonumber \\
		&= \frac{\partial\langle H_0 \rangle}{\partial\alpha} \dot{\alpha} + \frac{\partial\langle H_0 \rangle}{\partial\beta} \dot{\beta} + \frac{\partial\langle H_0 \rangle}{\partial\alpha^*} \dot{\alpha}^* + \frac{\partial\langle H_0 \rangle}{\partial\beta^*} \dot{\beta}^*, \label{eq:dHdt}
\end{align}
where we used a dot to indicate a time derivative. Let us find the partial derivatives of $\langle H_0 \rangle$ and then the time derivatives of $\alpha$, $\beta$, and their conjugates. First, we differentiate \eqref{eq:expected-H} to get
\begin{equation}
	\label{eq:H-partials}
    \begin{aligned}
	\frac{\partial\langle H_0 \rangle}{\partial \alpha}
		&= a\alpha^* + b^*\beta^*, \\
	\frac{\partial\langle H_0 \rangle}{\partial \beta}
		&= b\alpha^* + d\beta^*, \\
	\frac{\partial\langle H_0 \rangle}{\partial \alpha^*}
		&= a\alpha + b\beta, \\
	\frac{\partial\langle H_0 \rangle}{\partial \beta^*}
		&= b^*\alpha + d\beta.
    \end{aligned}
\end{equation}
Then, to find $\dot{\alpha}$, $\dot{\beta}$, $\dot{\alpha}^*$, and $\dot{\beta}^*$ in \eqref{eq:dHdt}, we use Schr\"odinger's equation \eqref{eq:SE}:
\[ i \begin{pmatrix}
	\dot{\alpha} \\
	\dot{\beta} \\
\end{pmatrix} = \begin{pmatrix} a & b \\ b^* & d \end{pmatrix} \begin{pmatrix}
	\alpha \\
	\beta \\
\end{pmatrix}, \]
which multiplied out is
\begin{align*}
    i \dot{\alpha} &= a \alpha + b \beta, \\
    i \dot{\beta} &= b^* \alpha + d \beta.
\end{align*}
Solving for $\dot{\alpha}$ and $\dot{\beta}$ and taking the complex conjugate of each equation,
\begin{equation}
	\label{eq:alpha-beta-dot}
    \begin{aligned}
        &\dot{\alpha} = -i(a \alpha + b \beta), \\
        &\dot{\beta} = -i(b^* \alpha + d \beta), \\
        &\dot{\alpha}^* = i(a \alpha^* + b^* \beta^*), \\
        &\dot{\beta}^* = i(b \alpha^* + d \beta^*).
	\end{aligned}
\end{equation}
Plugging \eqref{eq:H-partials} and \eqref{eq:alpha-beta-dot} into \eqref{eq:dHdt}, we get
\begin{align}
	\frac{d\langle H_0 \rangle}{dt}
		&= (a\alpha^* + b^*\beta^*)[-i(a \alpha + b \beta)] + (b\alpha^* + d\beta^*)[-i(b^* \alpha + d \beta)] \nonumber \\
        &\quad + (a\alpha + b\beta)[i(a \alpha^* + b^* \beta^*)] + (b^*\alpha + d\beta)[i(b \alpha^* + d \beta^*)] \nonumber \\
		&= 0, \label{eq:dH0dt}
\end{align}
so we have proved that the expected value of the Hamiltonian is constant.

Our derivation also shows that there is a relationship between the derivatives of the expected value of the Hamiltonian and the time derivatives of the amplitudes. Comparing \eqref{eq:H-partials} and \eqref{eq:alpha-beta-dot},
\begin{equation}
	\label{eq:H-alpha-beta-dot}
    \begin{aligned}
	\frac{\partial\langle H_0 \rangle}{\partial \alpha}
		&= -i \dot{\alpha}^*, \\
	\frac{\partial\langle H_0 \rangle}{\partial \beta}
		&= -i \dot{\beta}^*, \\
	\frac{\partial\langle H_0 \rangle}{\partial \alpha^*}
		&= i \dot{\alpha}, \\
	\frac{\partial\langle H_0 \rangle}{\partial \beta^*}
		&= i \dot{\beta}.
    \end{aligned}
\end{equation}
Plugging these into \eqref{eq:dHdt}, we get a cleaner proof that the expected value of the Hamiltonian is conserved:
\begin{align*}
	\frac{d\langle H_0 \rangle}{dt} 
		&= -i \dot{\alpha}^* \dot{\alpha} - i \dot{\beta}^* \dot{\beta} + i \dot{\alpha} \dot{\alpha}^* + i \dot{\beta} \dot{\beta}^* \\
        &= -i |\alpha|^2 - i |\beta|^2 + i |\alpha|^2 + i |\beta|^2 \\
        &= 0.
\end{align*}

Of course, the cleanest proof that the expected value of this time-independent Hamiltonian is conserved follows from the fact that the time evolution multiplies any eigenvector $\psi_j$ of $H_0$ by the phase $e^{iE_jt}$.  This leaves the norm-squared of the projection of the state onto each eigenspace invariant, and also, \textit{a fortiori}, $\langle H_0\rangle$.  This proof fails, however, for the time-dependent evolutions we will consider in the next two sections.  This is why we gave the preceding argument, which can be translated into the nonlinear settings.


\section{\label{section:repulsive}Nonlinear Quantum Search with Repulsive Interactions}

Many-body quantum systems can evolve by effective nonlinearities. Bose-Einstein condensates are a quintessential example, and for pairwise contact interactions in the mean field limit, they evolve by the Gross-Pitaevskii equation, which is a Schr\"odinger-type equation with a cubic nonlinearity \eqref{eq:NLSE}. A nonlinear quantum search algorithm \cite{Wong3} can be constructed from this by taking $H_0 = -\gamma L - \ketbra{a}{a}$ \eqref{eq:H0} so that the effective Hamiltonian \eqref{eq:H(t)} becomes
\begin{equation}
    \label{eq:H(t)-search}
    H(t) = -\gamma L - \ketbra{a}{a} + \lambda \sum_{i=1}^N \left| \braket{i}{\psi(t)} \right|^2 \ketbra{i}{i}.
\end{equation}
When the nonlinearity coefficient $\lambda$ is positive, this physically corresponds to bosons with repulsive interactions, and when $\lambda$ is negative, the interactions are attractive. The sign and strength of $\lambda$ can be tuned, for example, by Feshbach resonance \cite{Roberts2001}. In this section, we consider $\lambda > 0$, and in the next section, we consider $\lambda < 0$.

As shown in \cite{Kahou2013}, for large $N$, when $\gamma$ takes a critical value of
\[ \gamma_c = \frac{2-\lambda}{2N} \]
and the nonlinearity coefficient $\lambda$ lies in the range $0 < \lambda < \lambda_c$, where
\[ \lambda_c = \frac{4}{2+\sqrt{N}}, \]
then the system evolves from the initial uniform superposition \eqref{eq:psi(0)} to the marked vertex $\ket{a}$ in time roughly
\[ t_* = \frac{\pi}{\sqrt{3}} \sqrt{N}. \]
This runtime is slower than the linear algorithm's $\pi\sqrt{N}/2$ \eqref{eq:t*-gamma} by a factor of $2/\sqrt{3} \approx 1.155$, \textit{i.e.}, it is slower by 15.5\%. As discussed in \cite{Kahou2013}, although this algorithm searches in the same $O(\sqrt{N})$ time as the linear quantum algorithm, but with a worse coefficient, it nonetheless shows that it is theoretically possible for a Bose-Einstein condensate with repulsive interactions evolving by an effective nonlinearity to implement a quantum search algorithm with an $O(\sqrt{N})$ scaling.

\begin{figure}
\begin{center}
    \includegraphics{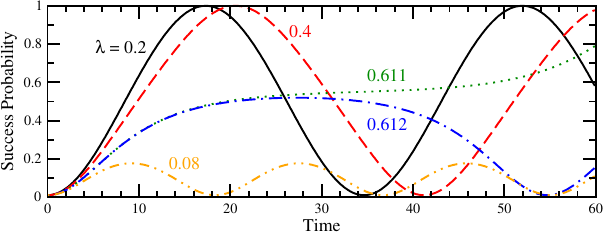}
	\vspace*{13pt}
    \fcaption{\label{fig:repulsive-n100}Success probability versus time for nonlinear quantum search on the complete graph with $N = 100$ vertices, $\gamma = (2-\lambda)/2N$, and various values of $\lambda$. The solid black curve is $\lambda = 0.2$, the dashed red curve is $\lambda = 0.6$, the dotted green curve is $\lambda = 0.611$, the dot-dashed blue curve is $\lambda = 0.612$, and the dot-dot-dashed orange curve is $\lambda = 0.8$.}
\end{center}
\end{figure}

Note from \cite{Kahou2013} that $\lambda_c$ is not a tight upper bound, \textit{i.e.}, there may exist some nonlinearity coefficient $\lambda > \lambda_c$ for which the success probability reaches 1, but eventually as $\lambda$ increases, the success probability stops reaching 1. This is shown in \fref{fig:repulsive-n100} for search on a complete graph of $N = 100$ vertices, for which $\lambda_c = 4/(2+\sqrt{100}) = 4/12 = 1/3 \approx 0.333$. The solid black curve is $\lambda = 0.2$, since $0 < 0.2 < 0.333$, the success probability must evolve to 1, as it does. The dashed red curve is $\lambda = 0.4$, and while it is greater than $\lambda_c = 0.333$, the success probability still reaches 1. The dotted green curve is $\lambda = 0.611$, and while the success probability struggles to grow, it does eventually reach 1 off the right of the graph. In contrast, the dot-dashed blue curve is $\lambda = 0.612$, and the success probability does not reach 1 anymore. A larger nonlinearity coefficient reaches a smaller and smaller success probability, such as the dot-dot-dashed orange curve with $\lambda = 0.08$. A tight upper bound on $\lambda$ such that the success probability reaches 1 remains an open question.

Since the effective Hamiltonian of the search algorithm depends on time, $\langle H(t) \rangle$ is generally not conserved. Instead, the expected value of $H_0$ plus half the nonlinear self-potential is conserved, \textit{i.e.}, $\langle H_0 + \frac{1}{2} \lambda|\psi|^2 \rangle$ is conserved, as this is the quantity that corresponds to the total energy \cite{Rogel-Salazar2013}. Conserved quantities for the Gross-Pitaevskii equation have been explored in some other contexts, including as a classical Hamiltonian \cite{Kahou2013}, the focusing cubic and quintic Gross-Pitaevskii hierarchies \cite{Chen2010}, cubic Gross-Pitaevskii hierarchy on $\mathbb{R}$ \cite{Mendelson2019}, and the one-dimensional Gross-Pitaevskii equation \cite{Koch2021,Koch2023}, but here we concentrate on conserved quantities related to the expected value of the effective Hamiltonian.

For instructional purposes, let us use the nonlinear search algorithm to explicitly show why $\langle H(t) \rangle = \langle H_0 + \lambda|\psi|^2 \rangle$ is not conserved, while $\langle H_0 + \frac{1}{2} \lambda|\psi|^2 \rangle$ is conserved. Even with the nonlinearity, as before, the system evolves in the same 2D subspace spanned by $\ket{a}$ and $\ket{b}$, so the state $\ket{\psi(t)}$ can be written as \eqref{eq:psi(t)-2D-general}. Now, the effective Hamiltonian \eqref{eq:H(t)-search} is
\begin{equation}
    \label{eq:H(t)-2D}
    H(t) = \begin{pmatrix}
	   \gamma(N-1) - 1 + \lambda |\alpha|^2 & -\gamma \sqrt{N-1} \\
	   -\gamma \sqrt{N-1} & \gamma + \frac{\lambda}{N-1} |\beta|^2 \\
    \end{pmatrix}.
\end{equation}
This takes the form
\[ H(t) = \begin{pmatrix} a + f|\alpha|^2 & b \\ b^* & d + g|\beta|^2 \end{pmatrix}, \]
where $a$, $b$, and $d$ are defined in \eqref{eq:abd}, and
\begin{equation}
    \label{eq:fg}
    f = \lambda \quad \text{and} \quad g = \frac{\lambda}{N-1}
\end{equation}
are real and time independent. Now, the expected value of the effective Hamiltonian is
\begin{align}
    \langle H(t) \rangle
		&= \langle \psi | H(t) | \psi \rangle \nonumber \\
		&= \begin{pmatrix} \alpha^* & \beta^* \end{pmatrix} \begin{pmatrix} a + f|\alpha|^2 & b \\ b^* & d + g|\beta|^2 \end{pmatrix} \begin{pmatrix} \alpha \\ \beta \end{pmatrix} \nonumber \\
		&= \begin{pmatrix} \alpha^* & \beta^* \end{pmatrix} \begin{pmatrix} a\alpha + b\beta + f|\alpha|^2\alpha \\ b^*\alpha + d\beta + g|\beta|^2\beta \end{pmatrix} \nonumber \\
		&= a\alpha^*\alpha + b\alpha^*\beta + f|\alpha|^4 + b^*\beta^*\alpha + d\beta^*\beta + g|\beta|^4. \label{eq:expected-H(t)}
\end{align}
Differentiating this with respect to time using the chain rule,
\begin{align}
    \frac{d\langle H(t) \rangle}{dt} 
		&= \frac{\partial\langle H(t) \rangle}{\partial\alpha} \frac{d\alpha}{dt} + \frac{\partial\langle H(t) \rangle}{\partial\beta} \frac{d\beta}{dt} + \frac{\partial\langle H(t) \rangle}{\partial\alpha^*} \frac{d\alpha^*}{dt} + \frac{\partial\langle H(t) \rangle}{\partial\beta^*} \frac{d\beta^*}{dt} \nonumber \\
		&= \frac{\partial\langle H(t) \rangle}{\partial\alpha} \dot{\alpha} + \frac{\partial\langle H(t) \rangle}{\partial\beta} \dot{\beta} + \frac{\partial\langle H(t) \rangle}{\partial\alpha^*} \dot{\alpha}^* + \frac{\partial\langle H(t) \rangle}{\partial\beta^*} \dot{\beta}^*, \label{eq:dH(t)dt}
\end{align}
The partial derivatives of the expected value of the effective Hamiltonian can be found by differentiating \eqref{eq:expected-H(t)}, noting that $|\alpha|^4 = (\alpha^*)^2 \alpha^2$ and $|\beta|^4 = (\beta^*)^2 \beta^2$:
\begin{equation}
	\label{eq:H(t)-partials}
    \begin{aligned}
	\frac{\partial\langle H(t) \rangle}{\partial \alpha}
		&= a\alpha^* + b^*\beta^* + 2 f |\alpha|^2 \alpha^*, \\
	\frac{\partial\langle H(t) \rangle}{\partial \beta}
		&= b\alpha^* + d\beta^* + 2 g |\beta|^2 \beta^*, \\
	\frac{\partial\langle H(t) \rangle}{\partial \alpha^*}
		&= a\alpha + b\beta + 2 f |\alpha|^2 \alpha, \\
	\frac{\partial\langle H(t) \rangle}{\partial \beta^*}
		&= b^*\alpha + d\beta + 2 g |\beta|^2 \beta.
    \end{aligned}
\end{equation}
To find $\dot{\alpha}$, $\dot{\beta}$, $\dot{\alpha}^*$, and $\dot{\beta}^*$ in \eqref{eq:dH(t)dt}, we use the cubic nonlinear Schr\"odinger equation \eqref{eq:NLSE},
\[ i \begin{pmatrix} \dot{\alpha} \\ \dot{\beta} \end{pmatrix} = \begin{pmatrix} a + f|\alpha|^2 & b \\ b^* & d + g|\beta|^2 \end{pmatrix} \begin{pmatrix} \alpha \\ \beta \end{pmatrix}, \]
which multiplied out is
\begin{align*}
	i\dot{\alpha} &= a\alpha + b\beta + f|\alpha|^2\alpha, \\
	i\dot{\beta} &= b^*\alpha + d \beta + g|\beta|^2\beta.
\end{align*}
Solving for $\dot{\alpha}$ and $\dot{\beta}$ and taking taking the complex conjugate of each equation,
\begin{equation}
	\label{eq:alpha-beta-dot-nonlinear}
	\begin{aligned}
		\dot{\alpha} &= -i(a\alpha + b\beta + f|\alpha|^2\alpha), \\
		\dot{\beta} &= -i(b^*\alpha + d \beta + g|\beta|^2\beta), \\
		\dot{\alpha}^* &= i(a\alpha^* + b^*\beta^* + f|\alpha|^2\alpha^*), \\
		\dot{\beta}^* &= i(b\alpha^* + d \beta^* + g|\beta|^2\beta^*).
	\end{aligned}
\end{equation}
Comparing \eqref{eq:H(t)-partials} and \eqref{eq:alpha-beta-dot-nonlinear}, we see that these are no longer proportional to partial derivatives of the expected value of the Hamiltonian, \textit{i.e.}, \eqref{eq:H-alpha-beta-dot} no longer holds. Then, \eqref{eq:dH(t)dt} is not equal to zero, so the expected value of the effective Hamiltonian is not conserved.

The reason why \eqref{eq:H(t)-partials} is not proportional to \eqref{eq:alpha-beta-dot-nonlinear} is the extra factor of 2 in the last term. We can eliminate this factor of 2 by dividing the nonlinear self-potential by 2. That is, we consider
\[ H_0 + \frac{1}{2} \lambda |\psi|^2 = \begin{pmatrix}
    \gamma(N-1) - 1 + \frac{1}{2} \lambda |\alpha|^2 & -\gamma \sqrt{N-1} \\
    -\gamma \sqrt{N-1} & \gamma + \frac{\lambda}{2(N-1)} |\beta|^2 \\
\end{pmatrix}, \]
which takes the form
\begin{equation}
    \label{eq:H2-2D}
    H_0 + \frac{1}{2} \lambda |\psi|^2 = \begin{pmatrix} a + \frac{1}{2} f|\alpha|^2 & b \\ b^* & d + \frac{1}{2} g|\beta|^2 \end{pmatrix},
\end{equation}
where $a$, $b$, and $d$ are defined in \eqref{eq:abd}, and $f$ and $g$ are defined in \eqref{eq:fg}. The expected value of this is
\begin{align}
    \left\langle H_0 + \frac{1}{2} \lambda |\psi|^2 \right\rangle
		&= \left\langle \psi \middle| H_0 + \frac{1}{2} \lambda |\psi|^2 \middle| \psi \right\rangle \nonumber \\
		&= \begin{pmatrix} \alpha^* & \beta^* \end{pmatrix} \begin{pmatrix} a + \frac{1}{2} f|\alpha|^2 & b \\ b^* & d + \frac{1}{2} g|\beta|^2 \end{pmatrix} \begin{pmatrix} \alpha \\ \beta \end{pmatrix} \nonumber \\
		&= \begin{pmatrix} \alpha^* & \beta^* \end{pmatrix} \begin{pmatrix} a\alpha + b\beta + \frac{1}{2} f|\alpha|^2\alpha \\ b^*\alpha + d\beta + \frac{1}{2} g|\beta|^2\beta \end{pmatrix} \nonumber \\
        &=a\alpha^*\alpha + b\alpha^*\beta + \frac{1}{2} f|\alpha|^4 + b\beta^*\alpha + d\beta^*\beta + \frac{1}{2} g|\beta|^4, \label{eq:expected-H2}
\end{align}
which has a time derivative of
\begin{align}
    \frac{d\langle H_0 + \frac{1}{2} \lambda |\psi|^2 \rangle}{dt} 
		&= \frac{\partial\langle H_0 + \frac{1}{2} \lambda |\psi|^2 \rangle}{\partial\alpha} \frac{d\alpha}{dt} + \frac{\partial\langle H_0 + \frac{1}{2} \lambda |\psi|^2 \rangle}{\partial\beta} \frac{d\beta}{dt} \nonumber \\
        &\quad + \frac{\partial\langle H_0 + \frac{1}{2} \lambda |\psi|^2 \rangle}{\partial\alpha^*} \frac{d\alpha^*}{dt} + \frac{\partial\langle H_0 + \frac{1}{2} \lambda |\psi|^2 \rangle}{\partial\beta^*} \frac{d\beta^*}{dt} \nonumber \\
		&= \frac{\partial\langle H_0 + \frac{1}{2} \lambda |\psi|^2 \rangle}{\partial\alpha} \dot{\alpha} + \frac{\partial\langle H_0 + \frac{1}{2} \lambda |\psi|^2 \rangle}{\partial\beta} \dot{\beta} \nonumber \\
        &\quad + \frac{\partial\langle H_0 + \frac{1}{2} \lambda |\psi|^2 \rangle}{\partial\alpha^*} \dot{\alpha}^* + \frac{\partial\langle H_0 + \frac{1}{2} \lambda |\psi|^2 \rangle}{\partial\beta^*} \dot{\beta}^*, \label{eq:dH2dt}
\end{align}
Now, when we take the partial derivatives of \eqref{eq:expected-H2}, the factors of 2 are eliminated:
\begin{align*}
	\frac{\partial\langle H_0 + \frac{1}{2} \lambda |\psi|^2 \rangle}{\partial \alpha}
		&= a\alpha^* + b\beta^* + f|\alpha|^2\alpha^*, \\
	\frac{\partial\langle H_0 + \frac{1}{2} \lambda |\psi|^2 \rangle}{\partial \beta}
		&= b\alpha^* + d\beta^* + g|\beta|^2\beta^*, \\
	\frac{\partial\langle H_0 + \frac{1}{2} \lambda |\psi|^2 \rangle}{\partial \alpha^*}
		&= a\alpha + b\beta + f|\alpha|^2\alpha, \\
	\frac{\partial\langle H_0 + \frac{1}{2} \lambda |\psi|^2 \rangle}{\partial \beta^*}
		&= b\alpha + d\beta + g|\beta|^2\beta.
\end{align*}
Comparing these with \eqref{eq:alpha-beta-dot-nonlinear}, these are exactly proportional to the time derivatives of $\alpha^*$, $\beta^*$, $\alpha$, and $\beta$, \textit{i.e.},
\begin{align*}
	\frac{\partial\langle H_0 + \frac{1}{2} \lambda |\psi|^2 \rangle}{\partial \alpha}
		&= -i \dot{\alpha}^*, \\
	\frac{\partial\langle H_0 + \frac{1}{2} \lambda |\psi|^2 \rangle}{\partial \beta}
		&= -i \dot{\beta}^*, \\
	\frac{\partial\langle H_0 + \frac{1}{2} \lambda |\psi|^2 \rangle}{\partial \alpha^*}
		&= i \dot{\alpha}, \\
	\frac{\partial\langle H_0 + \frac{1}{2} \lambda |\psi|^2 \rangle}{\partial \beta^*}
		&= i \dot{\beta}.
\end{align*}
Then, substituting into \eqref{eq:dH2dt}, we get
\begin{align*}
	\frac{d\langle H_0 + \frac{1}{2} \lambda |\psi|^2 \rangle}{dt} 
		&= -i \dot{\alpha}^* \dot{\alpha} -i \dot{\beta}^* \dot{\beta} + i \dot{\alpha} \dot{\alpha}^* + i \dot{\beta} \dot{\beta}^* \\
		&= 0.
\end{align*}
Thus, $\langle H_0 + \frac{1}{2} \lambda |\psi|^2 \rangle$ is conserved in the nonlinear quantum search algorithm with repulsive interactions.


\section{\label{section:attractive}Faster Nonlinear Quantum Search with Attractive Interactions}

\begin{figure}
\begin{center}
    \includegraphics{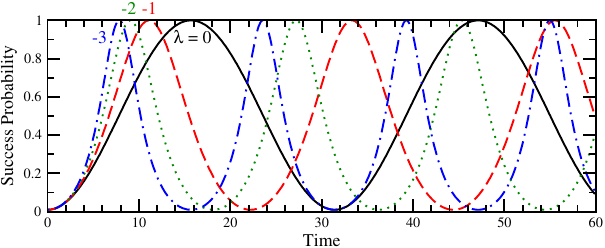}
	\vspace*{13pt}
    \fcaption{\label{fig:attractive-n100}Success probability versus time for nonlinear quantum search on the complete graph with $N = 100$ vertices, $\gamma = \gamma_c(t)$, and various values of $\lambda$. The solid black curve is $\lambda = 0$, the dashed red curve is $\lambda = -1$, the dotted green curve is $\lambda = -2$, and the dot-dashed blue curve is $\lambda = -3$.}
\end{center}
\end{figure}

In this section, we consider the nonlinear quantum search algorithm with negative values of $\lambda$, which corresponds to a Bose-Einstein condensate with attractive interactions. Intuitively, as the amplitude at the marked vertex builds up, the attractive interaction should draw additional amplitude, speeding up the algorithm. Indeed, as shown in \cite{Wong3}, when the jumping rate is the time-varying critical function
\begin{equation}
    \label{eq:gamma_c(t)}
    \gamma_c(t) = \frac{1}{N} \left[ 1 - \lambda \left( |\alpha|^2 - \frac{|\beta|^2}{N-1} \right) \right]
\end{equation}
[\textsl{which is determined independently of the marked vertex $|a\rangle$}, and whose time dependence comes from the amplitudes $\alpha$ and $\beta$ \eqref{eq:psi(t)-2D-general}], then the algorithm evolves from the uniform superposition over all $N$ vertices to the marked vertex in time
\[ t_* = \frac{1}{\sqrt{1-\lambda}} \frac{\pi}{2} \sqrt{N}. \]
This is demonstrated in \fref{fig:attractive-n100}, where we plot the evolution of the success probability for search on the complete graph of $N = 100$ vertices. The solid black, dashed red, dotted green, and dot-dashed blue curves correspond to $\lambda = 0$, $-1$, $-2$, and $-3$, and they reach success probabilities of 1 at time $t_* = (1/\sqrt{1-\lambda})(\pi/2)\sqrt{N} \approx 15.708, 11.107, 9.069, 7.854$, respectively, and so the more negative the nonlinearity coefficient, the faster the algorithm.

This speedup comes, however, at the expense of increasing the necessary time-measurement precision. In \fref{fig:attractive-n100}, as the magnitude of the nonlinearity coefficient increases, the peak in success probability becomes more narrow, which necessitates greater time-measurement precision to catch the peak. This was quantified in \cite{Wong3}, which showed that the width of the peak scales as $O(\sqrt{N}/(1-\lambda))$, and so the time-measurement precision is constant when $\lambda = \Theta(\sqrt{N})$, at which the runtime $t_* = O(N^{1/4})$, which is a quadratic improvement over the linear and nonlinear repulsive algorithms.

Since the effective Hamiltonian $H(t) = H_0 + \lambda |\psi|^2$ depends on time, its expected value is not conserved. $H_0 + \frac{1}{2} \lambda |\psi|^2$ takes the form given in \eqref{eq:H2-2D}, but now $a$, $b$, and $d$ depend on time because $\gamma$ varies with time. Then, the time-derivative of $\langle H_0 + \frac{1}{2} \lambda |\psi|^2 \rangle$ from \eqref{eq:dH2dt} will include additional terms involving time-derivatives of $a$, $b$, and $d$, and $\langle H_0 + \frac{1}{2} \lambda |\psi|^2 \rangle$ will not be conserved. Instead, let us show that another quantity is conserved.

First, we review from \cite{Wong3} that the critical function $\gamma_c(t)$ for the jumping rate causes $H(t)$ to be proportional to $H_0$. To begin,
\begin{align*}
    H(t)
        &= -\gamma_c(t) L - \ketbra{a}{a} + \lambda \sum_{i=1}^N \left| \braket{i}{\psi(t)} \right|^2 \ketbra{i}{i} \\
        &= -\gamma_c(t) L - \ketbra{a}{a} + \lambda \left| \braket{a}{\psi(t)} \right|^2 \ketbra{a}{a} + \lambda \sum_{i \ne a} \left| \braket{i}{\psi(t)} \right|^2 \ketbra{i}{i} \\
        &= -\gamma_c(t) L - \ketbra{a}{a} + \lambda \left| \alpha \right|^2 \ketbra{a}{a} + \lambda \frac{|\beta|^2}{N-1} \sum_{i \ne a} \ketbra{i}{i} \\
        &= -\gamma_c(t) L - \left( 1 - \lambda \left| \alpha \right|^2 \right) \ketbra{a}{a} + \lambda \frac{|\beta|^2}{N-1} \sum_{i \ne a} \ketbra{i}{i}.
\end{align*}
Now, recall that the jumping rate satisfies \eqref{eq:gamma_c(t)}. Multiplying both sides of \eqref{eq:gamma_c(t)} by $N$ and distributing $\lambda$ yields
\[ \gamma_c(t) N = 1 - \lambda |\alpha|^2 - \lambda \frac{|\beta|^2}{N-1}. \]
Rearranging,
\[ 1 - \lambda |\alpha|^2 = \gamma_c(t) N + \lambda \frac{|\beta|^2 }{N-1}. \]
Substituting this into our previous expression for $H(t)$,
\begin{align*}
    H(t)
        &= -\gamma_c(t) L - \left( \gamma_c(t) N + \lambda \frac{|\beta|^2}{N-1} \right) \ketbra{a}{a} + \lambda \frac{|\beta|^2}{N-1} \sum_{i \ne a} \ketbra{i}{i} \\
        &= -\gamma_c(t) L - \gamma_c(t) N \ketbra{a}{a} + \lambda \frac{|\beta|^2}{N-1} \sum_{i=1}^N \ketbra{i}{i} \\
        &= -\gamma_c(t) L - \gamma_c(t) N \ketbra{a}{a} + \lambda \frac{|\beta|^2}{N-1} \mathbb{I},
\end{align*}
where $\mathbb{I}$ is the identity matrix, which can be dropped since it corresponds to a rescaling of energy or a global, unobservable phase. Then,
\begin{align*}
    H(t)
        &= -\gamma_c(t) L - \gamma_c(t) N \ketbra{a}{a} \\
        &= \gamma_c(t) N \left( -\frac{1}{N} L - \ketbra{a}{a} \right) \\
        &= \gamma_c(t) N \left. H_0 \right|_{\gamma_c}.
\end{align*}
That is, the effective Hamiltonian for the nonlinear search algorithm with attractive interactions is a time-varying rescaling of the linear search algorithm's Hamiltonian evaluated at its critical jumping rate of $\gamma_c = 1/N$. Solving for the linear algorithm's Hamiltonian at its critical jumping rate,
\[ \left. H_0 \right|_{\gamma_c} = \frac{H(t)}{\gamma_c(t) N}. \]
Since $\langle H_0 \rangle$ is conserved for arbitrary constant $\gamma$, as we proved in \eqref{eq:dH0dt}, we have that
\[ \left\langle \frac{H(t)}{\gamma_c(t) N} \right\rangle = 0, \]
and we have proved a conserved quantity for the nonlinear search algorithm with attractive interactions.


\section{\label{section:conclusion}Conclusion}

We have reviewed three quantum search algorithms governed by continuous-time quantum walks. The first used a linear quantum walk, and we derived the evolution of the system for arbitrary jumping rates. We used elementary calculus to prove that there is a critical jumping rate $\gamma_c$ that causes the success probability to reach $1$ in time $\pi\sqrt{N}/2$. We showed that the expected value of the Hamiltonian $H_0$ is conserved, which is an example of the fact that all time-independent Hamiltonians have constant expected values. Second, we considered a nonlinear quantum walk with repulsive interactions, whose effective Hamiltonian was $H(t) = H_0 + \lambda |\psi|^2$ with $\lambda > 0$. For a range of possible critical jumping rates, the success probability reaches $1$ in time $\pi\sqrt{N/3}$, which is slower than the linear algorithm by a constant factor, and we showed that the expected value of $H_0 + \frac{1}{2} \lambda |\psi|^2$ is conserved. Third, for the nonlinear quantum walk with attractive interactions, the effective Hamiltonian has $\lambda < 0$, and when the jumping rate takes a critical function $\gamma_c(t)$, the success probability reaches $1$ in time $\pi\sqrt{N/(1-\lambda)}/2$, which is faster than the other algorithms for all $\lambda < 0$. The jumping rate varies such that the effective Hamiltonian $H(t)$ is equal to a time-dependent rescaling of the linear algorithm's Hamiltonian $H_0$ evaluated at its critical jumping rate $\gamma_c$, namely $H(t) = \gamma_c(t) N \left. H_0 \right|_{\gamma_c}$, and since the expected value of the time-independent linear Hamiltonian $H_0$ is conserved for an arbitrary constant jumping rate, the expected value of $H(t)/[\gamma_c(t) N]$ is conserved for the third algorithm. These results provide a perspective on quantum walk search algorithms through the lens of conservation laws and many-body quantum theory.


\nonumsection{Acknowledgements}
\noindent
This material is based upon work supported in part by the National Science Foundation EPSCoR Cooperative Agreement OIA-2044049, Nebraska’s EQUATE collaboration. Any opinions, findings, and conclusions or recommendations expressed in this material are those of the author(s) and do not necessarily reflect the views of the National Science Foundation.


\nonumsection{References}
\noindent
\bibliography{refs}
\bibliographystyle{qic}

\end{document}